\documentclass[prl, twocolumn, amsmath,floatfix, unsortedaddress, superscriptaddress]{revtex4-1}

\usepackage{graphicx}
\usepackage{epstopdf}
\usepackage{natbib}
\usepackage{xcolor}

\newcommand{\sciexp}[2]{{#1}\ensuremath{\,\times\,10^{#2}}}

\newcommand{\DHe}{D$^3$He}

\newcommand{\PPPL}{Princeton Plasma Physics Laboratory, Princeton, NJ 08543, USA}
\newcommand{\PAS}{Department of Astrophysical Sciences, Princeton University, Princeton, NJ 08544, USA}
\newcommand{\UMich}{Center for Ultrafast Optical Science, University of Michigan, Ann Arbor, MI 48109}

\newcommand{\LLE}{Laboratory for Laser Energetics, University of Rochester, Rochester, New York 14623, USA}

\newcommand{\CU}{University of Colorado, Boulder, CO 80309, USA}
\newcommand{\LLNL}{Lawrence Livermore National Laboratory, Livermore, CA, USA}
\newcommand{\MIT}{Plasma Science and Fusion Center, Massachusetts Institute of Technology, Cambridge, MA 02139, USA}

\begin{document}

\begin{abstract}

Fast magnetic reconnection was observed between 
magnetized laser-produced plasmas at the National Ignition Facility.
Two highly-elongated plasma plumes were produced by tiling
two rows of lasers, with magnetic fields generated in each plume
by the Biermann battery effect.  
Detailed magnetic field observations, obtained from proton radiography using 
a \DHe{} capsule implosion, reveal reconnection occurring in an extended, quasi-1D 
current sheet with large aspect ratio $\sim 100$.
The 1-D geometry allowed a rigorous and unique reconstruction of
the magnetic field, which showed a 
reconnection current sheet that thinned down to a half-width close to the 
electron gyro-scale.  Despite the large aspect ratio,
a large fraction of the magnetic flux reconnected,
suggesting fast reconnection supported
by the non-gyrotropic electron pressure tensor.
\end{abstract}

\title{Fast magnetic reconnection in highly-extended current sheets at the National Ignition Facility}
\author{W. Fox}
\email{wfox@pppl.gov}
\affiliation{\PPPL{}}
\affiliation{\PAS{}}
\author{D.B. Schaeffer}
\affiliation{\PAS{}}
\author{M.J. Rosenberg}
\affiliation{\LLE{}}
\author{G. Fiksel}
\affiliation{\UMich{}}
\author{J. Matteucci}
\affiliation{\PAS{}}
\author{H.-S. Park}
\affiliation{\LLNL{}}
\author{A. F. A. Bott}
\affiliation{\PAS{}}
\author{K. Lezhnin}
\affiliation{\PAS{}}
\author{A. Bhattacharjee}
\affiliation{\PPPL{}}
\affiliation{\PAS{}}
\author {D. Kalantar}
\author{B.A. Remington}
\affiliation{\LLNL{}}
\author{D. Uzdensky}
\affiliation{\CU{}}
\author{C.K. Li}
\author{F.H. S\'eguin}
\affiliation{\MIT{}}
\author{S.X. Hu}
\affiliation{\LLE{}}

\date{\today}

\maketitle

Magnetic reconnection enables the rapid conversion of magnetic field energy 
to bulk plasma kinetic energy and energized particle populations in plasmas ranging from 
laboratory to astrophysical environments \cite{YamadaRMP2010, Kulsrud2005}.
Important aspects of this phenomenon that are not yet understood include the structure and dynamics of 
magnetic reconnection current sheets, and particularly the rate of reconnection, in highly-extended sheets 
much longer than intrinsic plasma kinetic scales such as the particle skin-depth and gyro-radii \cite{JiPoP2011}.  
Laser-produced plasmas, in which magnetic fields are
generated in expanding plumes by the Biermann battery 
effect \cite{NilsonPRL2006, LiPRL2007b, RosenbergPRL2015},
offer a new experimental platform for reconnection studies with a 
large scale separation between the global system size
and kinetic plasma scales.
These experiments are  
further complementary to low-temperature discharge-type plasmas \cite{RenPRL2005,FoxPRL2017}
by observing current sheets at high-$\beta$ ($\beta = 2 \mu_0 n T / B^2$); 
reconnection at high-$\beta$, while it may not cause significant bulk plasma heating, may 
nevertheless be important
for modifying global magnetic topology and transport in astrophysical systems such as accretion disks \cite{QuataertAN2003}, and may
energize superthermal particles \cite{MatsumotoScience2015}.
Laboratory experiments also complement studies with spacecraft \cite{BurchScience2016},
by enabling observation 2-D and 3-D reconnection dynamics spanning global and kinetic scales.

Detailed current sheet observations can reveal important
aspects of reconnection in the high-$\beta$ regime.
The current sheet width 
determines the growth rates of current sheet instabilities in fast reconnection models based on tearing or plasmoid instabilities \cite{LoureiroPoP2007, BhattacharjeePoP2009, JiPoP2011}.
Furthermore, the width controls various mechanisms for breaking magnetic field lines.
Momentum transport (the electron pressure tensor term in Ohm's law),
has been observed in collisionless reconnection simulations as the mechanism to
 break field lines \cite{CaiPoP1997,HessePoP1999}.
  The physical mechanism \cite{Kulsrud2005} is that 
 thermal motion allows electrons to transit out of the electron diffusion region before they 
 can be accelerated by the reconnection electric field.
Recent work at the Magnetic Reconnection eXperiment (MRX)
studied electron diffusion regions at $\beta \sim 1$, though
found the diffusion region was too broad for momentum transport 
to play a role \cite{JiGRL2008, FoxPRL2017}, instead suggesting anomalous resistivity supplied by instabilities.
Electron diffusion regions are also observed in
space plasmas, and indeed resolving this kinetic scale is a primary goal of the Magnetosphere 
Multiscale (MMS) Mission \cite{BurchScience2016, PhanNature2018}.

\begin{figure*}
\includegraphics[width=6.5in]{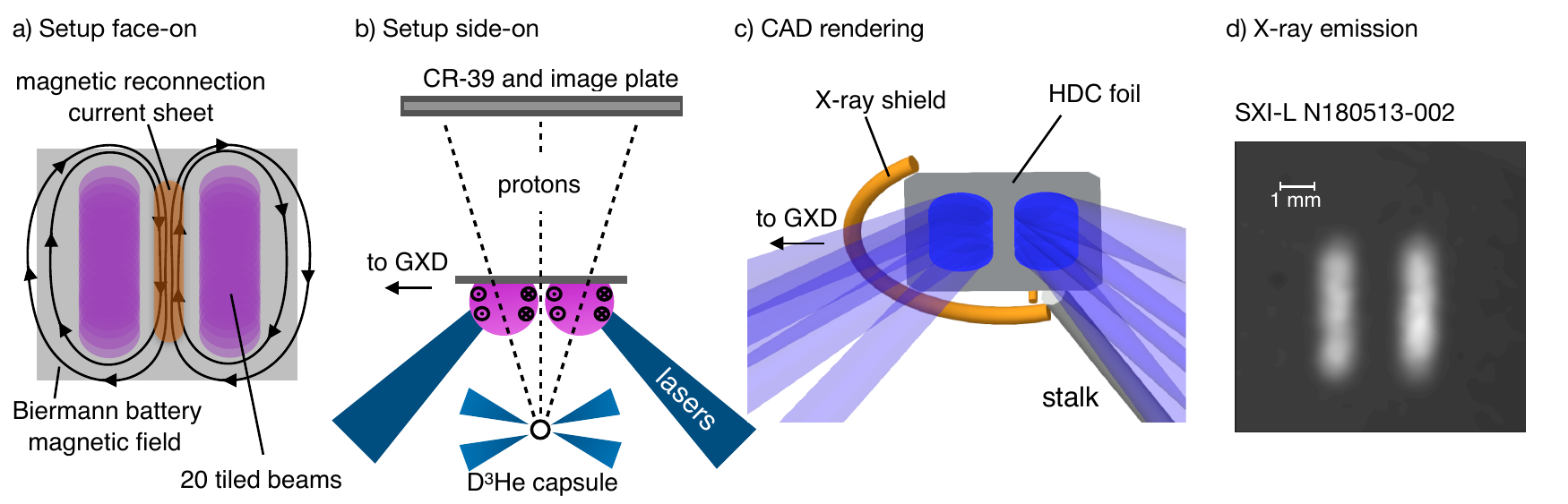}
\caption{Experimental setup.  a) Setup face-on, showing tiled laser foci producing two extended
plasma plumes, self-generated magnetic fields, and formation of an extended 1-D current sheet.  b) Side-on view, showing proton radiography geometry from
a \DHe{}  backlighter.  c) CAD rendering of the target, oblique view.\
d) Time-integrated soft x-ray emission showing two elongated plumes.}
\label{Fig_Setup}
\end{figure*}

In this Letter, we present results from experiments at the National Ignition Facility (NIF) to study 
reconnection in highly-extended current sheets much larger than
intrinsic plasma scales.  The magnetic fields 
are self-generated in two neighboring plasma plumes by the Biermann battery ($\nabla n \times \nabla T$) 
effect \cite{YatesPRL1982, MatteucciPRL2018}.  
On NIF, a large number of lasers are tiled to produce two
highly-elongated parallel plasma plumes (Fig.~\ref{Fig_Setup}), which 
merge by quasi-1-D inflows, in contrast 
to previous experiments which interacted two spherical expanding plumes.
The magnetic fields are reconstructed from proton radiography data, which reveals reconnection occurring in a highly extended
sheet, with a current sheet length $L_y = $~6~mm, much longer than
the half-width $\delta = 28 \pm 5$~$\mu$m, for an aspect ratio $A = L_y / 2 \delta \gtrsim 100$. 
A significant fraction of the magnetic flux is observed to reconnect, indicating a fast
reconnection mechanism. 
The current sheet half-width $\delta$  is close to the electron gyro-radius $\rho_e \sim~15$~$\mu$m
at these plasma conditions, suggesting that
momentum transport (the pressure tensor effect \cite{HessePoP1999}) drives fast reconnection.

Figure \ref{Fig_Setup} shows the experimental setup. 
 Extending the techniques of previous laser-driven reconnection experiments~\cite{NilsonPRL2006,LiPRL2007b,FikselPRL2014,RosenbergNatComm2015, RosenbergPRL2015}, 
 two sets of beams were focused to generate a pair of elongated plasma plumes.
In the NIF ``MagRecon'' platform, 40 beams are tiled evenly onto two parallel 4~$\times$~1~mm$^2$ footprints on a thin (15~$\mu$m) 5.5$\times$6~mm$^2$
high-density-carbon (HDC) foil to generate two colliding plasma plumes (Fig.~\ref{Fig_Setup}a). 
The lateral separation of the laser foci between the two footprints was 2.4~mm.  
 In these experiments, the total laser energy per plume was 2~kJ in 0.6~ns, for 
an on-target laser intensity of $I_L = 1 \times10^{14}$~W/cm$^2$ at  351~nm wavelength. 
The platform uses beams from the bottom outer beam cones (45--50$^{\circ}$) on NIF, with standard NIF phase plates which all project to smooth circular foci with
a 1.2~mm-diameter profile in the horizontal plane of the target.
A \DHe{} backlighter capsule for proton deflectometry was 
mounted below the target (Fig.~\ref{Fig_Setup}b), with the 
CR-39 detector package at the pole.   Gated X-ray Detectors (GXD) \cite{OertelRSI2006} 
were fielded in the plane of the target. 
Figure~\ref{Fig_Setup}d shows time-integrated soft x-ray emission from the target 
showing two uniform extended plumes.

Experiments observed the line-integrated magnetic fields using proton deflectometry.
The proton backlighter used an 860~$\mu$m-diameter glass capsule target filled with a
\DHe{} mixture (6~atm~D$_2$, 12~atm~$^3$He).   The capsule was imploded by 64 NIF beams, 
using a 1.3~ns square pulse shape, 0.9~kJ/beam, and standard NIF phase plates.  Upon implosion,
\DHe{} fusion reactions produced a burst of energized protons with a narrow spectrum peaked at 14.95~MeV, as recorded on nuclear detectors \cite{SeguinRSI2003}.
 The capsule was mounted 20~mm below the target foil,  while the
CR-39~detector stack was fielded 215~mm opposite the target, for a geometrical magnification $M = 11.75$.
An image plate (IP) was fielded behind the CR-39 foils in the stack to obtain a point-projection x-ray
image of the target region; comparing the IP to photographs of the as-built targets 
allowed registration of the proton radiographs against the physical targets.

\begin{figure}
\includegraphics[width=3in]{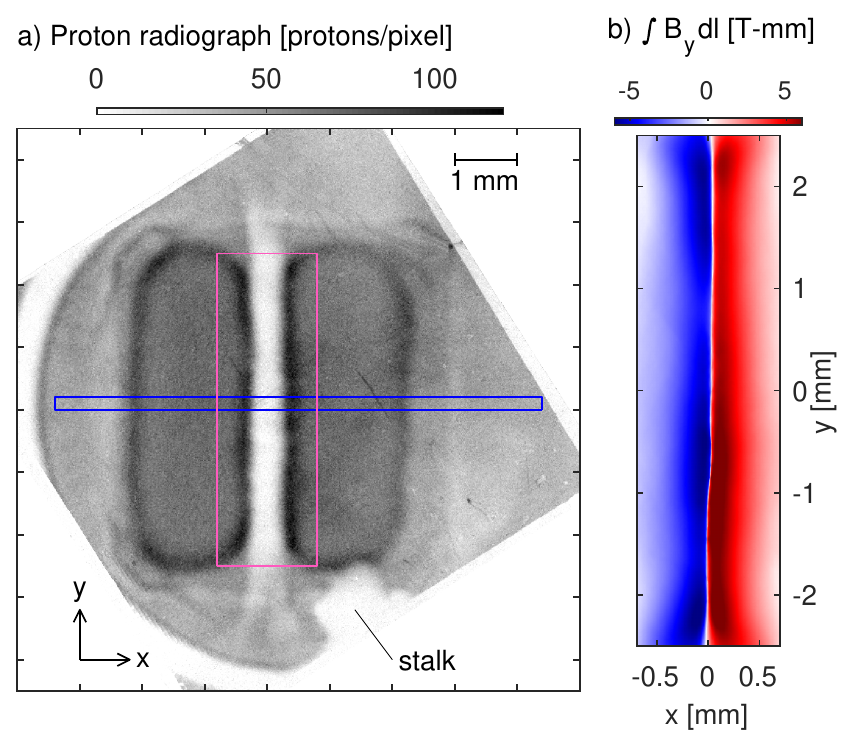}
\caption{a) Proton radiograph for two colliding plumes at  $t=3.0$~ns.  
The magenta box in analyzed in (b), which shows the 2-D reconstruction
of $\int B_y\,d\ell$ over the current sheet.  The blue box is analyzed in Fig.~\ref{Fig_Prad_1d}.  
Units correspond to the target plane accounting for magnification.}  
\label{Fig_Prad}
\end{figure}

Figure~\ref{Fig_Prad} shows proton radiography data  and reconstructed magnetic fields obtained from these experiments at time 
$t=3.0$~ns after the start of the main plume beam, accounting for the backlighter bang-time and proton time-of-flight.
As the protons stream through the interacting
 plumes, they pick up small deflections, predominantly from $B$ fields \cite{PetrassoPRL2009},
 which causes fluence variation on the CR-39 (Fig.~\ref{Fig_Prad}a.)
 Two high-fluence (dark), elongated “race-tracks” are immediately apparent which correspond to a focusing pileup of protons interior to the peak Biermann magnetic fields surrounding each plume. 
Where the two plumes interact, a region of extreme proton depletion is observed.  This is consistent with
a current sheet structure causing outward deflection of the probe protons.
The current sheet is quasi-1-D ($\partial/\partial y \ll \partial/\partial x$).    Figure~\ref{Fig_Prad}b shows the reconstructed magnetic fields
in the 2-D region containing the current sheet (the magenta box in Fig.~\ref{Fig_Prad}a), using the 2-D inversion algorithm PROBLEM \cite{BottJPP2017}.
The current sheet half-width $\delta$ varies along the sheet and ranges from 15-30~$\mu$m, much
thinner than the apparent width in the fluence image. Meanwhile the 
length of the sheet is $L_y \approx$~6~mm,
for an overall aspect ratio  $A = L_y/2\delta \gtrsim 100$.
Some knots and modulations of the proton fluence are also observed at the current sheet, as well as regions along the sheet which
have reconnected more than their neighbors.

The experimental quasi-1-D geometry allows a rigorous validation of the magnetic field reconstruction ({Fig.~\ref{Fig_Prad_1d}).
Fig.~\ref{Fig_Prad_1d}a shows, in the magenta curve, the observed proton fluence over the entire 1-D trace from the blue region of Fig.~\ref{Fig_Prad}a.   This lineout
is analyzed using a separate, fast 1-D reconstruction code PRADICAMENT \footnote{See extended material}.
Reconstructing magnetic fields from the fluence variations requires both the final proton 
fluence on the detector and the initial ``undisturbed'' proton fluence.  
Since there is usually no separate measurement 
of the undisturbed proton fluence along the same line of sight,
the undisturbed fluence must be constrained by other means, and indeed minor changes to the
undisturbed profile leads to large changes in the inferred $B$ field.
To constrain the possible $B$-field profiles, we 
first define a family of undisturbed fluence profiles, which
are assumed to be parabolic as a lowest-order shape,  and 
use PRADICAMENT to solve for the 
$B$ field for each fluence profile.
We then accept the $B$-field profiles (and associated undisturbed fluence profiles) which satisfy: 
(1) the $B$ fields do not reverse sign after their initial peaks, and approach zero toward the edge of the measurement region 
\cite{PetrassoPRL2009}; and (2), the $B$ fields are consistent with  target-edge magnetic field observations within error bars.   
These target-edge $B$ fields are displayed as the
two data points with error bars at $x = -2.5$ and +3~mm in Fig.~\ref{Fig_Prad_1d}b; they were determined by comparing 
the position of the target boundary in the proton radiography track-diameter image (not shown) to the physical target.  
The inward shift indicated a finite $B$ field at each target edge, with error bars arising from alignment and registration uncertainties.
The final set of satisfactory undisturbed fluence profiles is shown as the blue band in Fig.~\ref{Fig_Prad_1d}a, and the associated
$B$-field profiles as the blue band in Fig.~\ref{Fig_Prad_1d}b.  
The analysis indicates that the undisturbed proton fluence 
was not uniform over the measurement region by $\sim25\%$, 
a non-uniformity much larger than the $\cos^2 \theta$ factor from equal protons per unit angle 
(possibly resulting from electromagnetic fields in the corona of the backlighter target \cite{SeguinPoP2012}).
These fluence profiles were then used as input for the PROBLEM 2-D reconstruction, and we obtained good agreement between
the results in the central region.
The magnetic field crosses zero at  the two laser foci, $x=\pm 1.2$~mm, within error bars, which is expected 
since the Biermann generation should reach a minimum at those locations.

\begin{figure}
\includegraphics[width=3.375in]{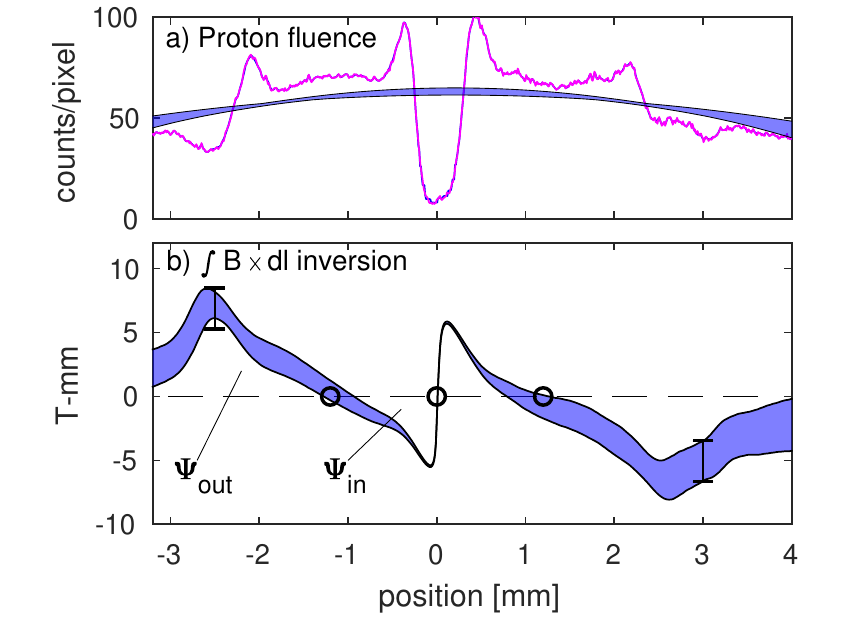}
\caption{Detailed 1-D magnetic field reconstructions.  
a) Magenta curve: proton fluence line-out over the blue region of Fig.~\ref{Fig_Prad}a;
Blue band: undisturbed fluence profiles utilized in the reconstruction.
b) $\int B_y \times d\ell$ reconstructed from the proton fluence, where the blue band corresponds to
magnetic reconstructions for the entire range of incident proton-fluence profiles in (a).
$\Psi$ labels regions for evaluating magnetic flux $\Psi = \int dx \int B_y \times d\ell$ to obtain inner and outer fluxes.}
\label{Fig_Prad_1d}
\end{figure}

\begin{figure*}
	\centering
	\includegraphics[width=6in]{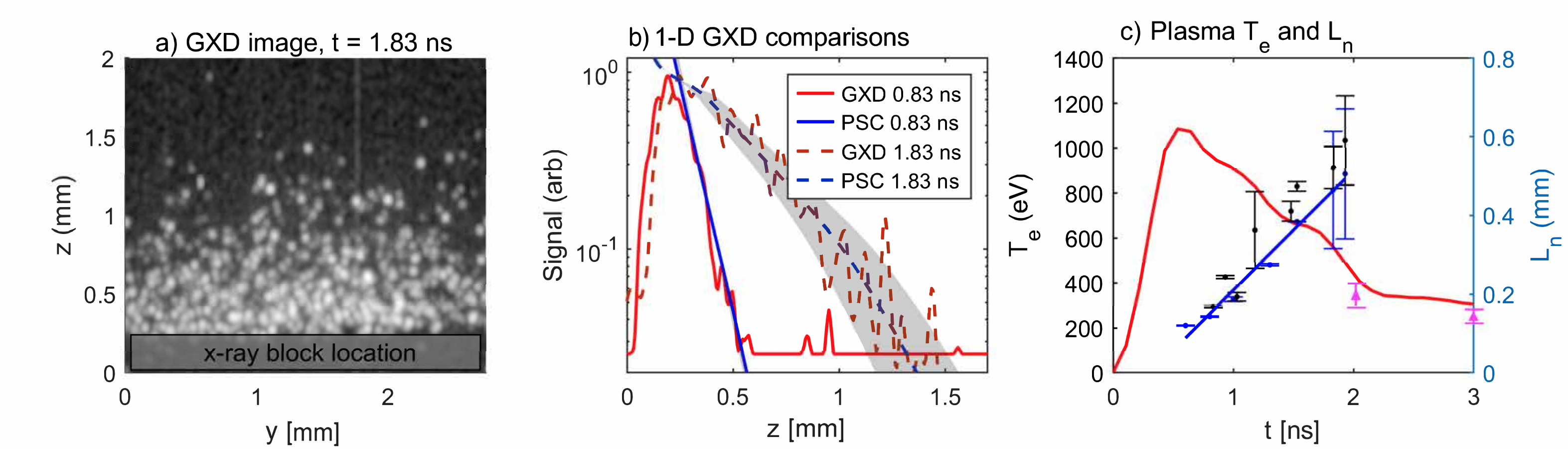}
	\caption{Analysis of experimental GXD with comparison to PSC. (a) 2-D GXD pinhole image obtained at 1.83 ns after laser pulse turns on.  The laser plasma expands upwards from the target at $z \approx 0$.  (b) Comparison of two vertical profiles of GXD (integrated across $y$) in red with synthetic signal obtained from PSC, shown in blue.  The gray shading shows varying of PSC heating by $\pm$ 20\%.
(c) Temperature measured within the current sheet at $z=+1$~mm from PSC shown in red vs. time.
Magenta data points are obtained directly from GXD x-ray ratios through 3- and 6-$\mu$m 
Al filters. Black: plume scale lengths $L_n = \langle n / (dn/dz) \rangle $ from GXD in black,
compared to scale lengths obtained from PSC in blue.}
\label{Fig_GXD}
\end{figure*}

\begin{figure}
\includegraphics[width=3.375in]{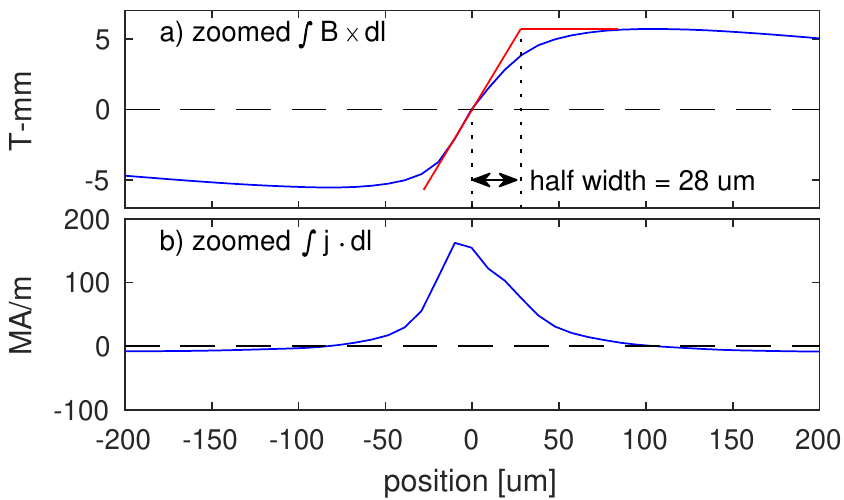}
\caption{Zoomed 1-D magnetic field reconstructions.  a) Zoom-in on the narrow current sheet showing
the reversal with half-width 28~$\mu$m. b) Profile of line-integrated current $\int j_z d\ell$.}
\label{Fig_Prad_1d_zoomed}
\end{figure}

Further plasma parameters were provided by the GXD which acquires gated pinhole images
of plasma Bremsstrahlung x-ray emission through various filters.
The GXD view was from nearly side-on in the equatorial plane of the target (Fig.~\ref{Fig_Setup}c) 
and observed the vertical plume expansion versus time (rather than current sheet structures).  
Fig.~\ref{Fig_GXD}a shows a typical side-on pinhole image showing the ablating plasma.
Since the profile evolution is strongly dependent on the plasma temperature and density, 
we used it to constrain a PSC particle-in-cell ablation simulation 
of two neighboring plumes \cite{FoxPoP2018}, by tuning the laser 
deposition to match the spatial evolution of Bremsstrahlung emission to
the observed GXD profiles.
The PSC simulation includes Biermann-battery magnetic field generation and the 
formation of a current sheet \cite{MatteucciPRL2018}.
Figure~\ref{Fig_GXD}b shows the Bremsstrahlung emission profiles at two times, where the gray shading shows a sensitivity test for 
PSC in which the heat deposition was modified by $\pm 20$\%.
The simulation showed that the plasma temperatures 
in the current sheet varied from 1000~eV at $t=0.6$~ns when the laser turns off to 300~eV at 3~ns (Fig.~\ref{Fig_GXD}c).
The results are consistent with plasma temperature inferred from 
GXD ratios through 3- and 6-$\mu$m Al filters (magenta points).
The PSC simulations additionally show densities in the current sheet ranging from 
1--3$\times 10^{25}$~m$^{-3}$ over the first 3~ns, for typical plasma $\beta$ of $\sim$~10--50. 

Connecting these observations provides valuable new data on reconnection in high-$\beta$ plasmas.
The $B_y$ profile can be integrated along the inflow direction ($x$), which yields the total magnetic flux contained in each bubble $\Psi = \int dx \int B_y d\ell$ (Fig.~\ref{Fig_Prad_1d}b).
A larger total magnetic flux is observed on the outer (non-reconnecting) side $\Psi_{out} = 5.5$--15~T-mm$^2$ of the bubbles than on the inside $\Psi_{in} = 2.5 \pm 0.3$~T-mm$^2$,
which confirms that at this time a significant amount of flux has reconnected or dissipated 
in the current sheet 
\footnote{PSC simulations matched to the plasma conditions showed that Biermann battery magnetic field generation was nearly equal between the inboard and outboard sides}. 
The minimum amount of flux reconnected ($\Delta \Psi = \Psi_{out} - \Psi_{in}$) is 
well constrained to $> 3$~T-mm$^2$, which is larger than the remaining inboard flux and indicates a minimum reconnection fraction of ~$\sim 60\%$.  
Using $\langle \int E_z d\ell \rangle \equiv \Delta \Psi / \Delta t$  by Faraday's law, we estimate a lower-bound,
time-averaged, vertically-integrated reconnection electric field $\langle \int E_z d\ell \rangle \sim$~1300~V over 3.0~ns.  
The average reconnection inflow speed of the reconnecting flux is 
correspondingly $\langle V_{in} \rangle \equiv \int E_zd\ell / \int B_y d\ell \sim$~\sciexp{2}{5}~m/s.
Using a plasma density of $10^{25}$~m$^{-3}$, this average inflow is super-Alfvenic, with $V_A \sim$~\sciexp{5}{4}~m/s, 
indicating a very strongly-driven reconnection.

In addition to inference of high magnetic reconnection rates, we obtain detailed observation of the 
reconnection current sheet.  
Figure~\ref{Fig_Prad_1d_zoomed}a zooms in on the current sheet, showing the
half-width $\delta = B_{up} / |dB/dx| = 28 \pm 5 \mu$m.
The uncertainty in the current sheet width is set by uncertainty in the CR-39 background level,
which was determined by analyzing CR-39 track noise under the border frame.
The current sheet has thinned to nearly the 
electron gyroradius, which is $\sim 15~\mu$m for $B =5$~T (for a 1~mm integration length typically shown in simulations) 
and $T_e = 400$~eV.

Using the magnetic profiles, we obtain the profile of 
line-integrated currents $\int j_z d\ell$ which is strongly peaked in the current
sheet at 160~MA/m (Fig.~\ref{Fig_Prad_1d_zoomed}b). 
This implies highly anomalous reconnection rates not supported by classical Spitzer resistivity,
which can only support $\sim \eta_{Sp} \int j_z d\ell \sim 50$V at 400~eV and Z = 6,  a
factor $\sim25$ below the 
reconnection rate of $\langle \int E_z d\ell \rangle \sim$~1300~V calculated above.
We note similar anomalous reconnection rates ($E_{rec} \gg \eta_{Sp} j$) were  observed at MRX
(under much different plasma conditions, and at $\beta \sim 1$) \cite{KuritsynPoP2006}.
We emphasize that the present data indicate only an averaged reconnection rate,
and at this time it is not clear that the instantaneous rate is still fast \cite{RosenbergPRL2015, LezhninPoP2018}.   
Nevertheless, the thinness of the current sheet suggests the importance of 
momentum transport (pressure tensor) in obtaining fast reconnection:
in the model of Hesse \cite{HessePoP1999}, the pressure tensor provides 
a reconnection electric field $E_{rec} \approx (1/e) \sqrt{2 m_e T_e} \;\partial{V_e}/\partial{x}$,
where $V_e$ is the electron inflow to the reconnection layer.  
Taking inflow speeds at the average global inflow, $V_{in}  \sim 200$~km/s, and 
$\partial / \partial x$ corresponding to the observed half-width $\sim 1/\delta \sim 1/25~\mu$m,  $T_e =$~400--1000~eV,
and an integration length $L \sim 1$~mm,  we estimate $\int E d\ell \propto \sqrt{T_e} V_{in} L / \delta \sim$~600--900~V.  While a rough
estimate, this is remarkably comparable to the observed rates, and argues for the importance of dissipation by pressure tensor 
in high-$\beta$ reconnection.
This effect was observed in previous simulations of reconnection in laser plasmas  \cite{FoxPRL2011} including 
recent full 3-D simulations \cite{MatteucciPoP2019}.   Future experiments measuring the instantaneous
reconnection rates and spatial profiles of plasma parameters in the current sheet can test the theory in detail.

To conclude, we have presented detailed magnetic reconnection observations 
at the NIF at high plasma $\beta$.  Tiling lines of  beams allows the generation of highly-extended reconnection sheets, driven
by well-controlled 1-D inflows.  
Validated magnetic field reconstructions using the proton data reveal 
a highly-extended, high-aspect ratio sheet with $A = L/2\delta \gtrsim 100$, with the current
sheet half-width $\delta$ narrowed down to the electron gyro-scale.  The  inferred high reconnection rates suggest the 
importance of electron momentum transport due to the narrowing of the current sheet to the electron gyro-scale.

\begin{acknowledgments}
We thank the NIF Discovery Science Program for facilitating these experiments. Support for these experiments has been provided by the U.S. DOE, Office of Fusion Energy Sciences under FWP SW1626 FES.  This work was performed under the auspices of the U.S. DOE by Lawrence Livermore National Laboratory under contract DE-AC52-07NA27344
and under DOE Award Nos. DE-SC-0016249, and DE-NA-0003856, 
and Field Work Proposal No. 4507 under DOE Contract No. DE-AC02-09CH11466.
Simulations were conducted on the Titan supercomputer at the Oak Ridge Leadership Computing Facility, supported by the Office of Science of the DOE under Contract No. DE-AC05-00OR22725.
\end{acknowledgments}
%

\begin{thebibliography}{35}
\makeatletter
\providecommand \@ifxundefined [1]{%
 \@ifx{#1\undefined}
}%
\providecommand \@ifnum [1]{%
 \ifnum #1\expandafter \@firstoftwo
 \else \expandafter \@secondoftwo
 \fi
}%
\providecommand \@ifx [1]{%
 \ifx #1\expandafter \@firstoftwo
 \else \expandafter \@secondoftwo
 \fi
}%
\providecommand \natexlab [1]{#1}%
\providecommand \enquote  [1]{``#1''}%
\providecommand \bibnamefont  [1]{#1}%
\providecommand \bibfnamefont [1]{#1}%
\providecommand \citenamefont [1]{#1}%
\providecommand \href@noop [0]{\@secondoftwo}%
\providecommand \href [0]{\begingroup \@sanitize@url \@href}%
\providecommand \@href[1]{\@@startlink{#1}\@@href}%
\providecommand \@@href[1]{\endgroup#1\@@endlink}%
\providecommand \@sanitize@url [0]{\catcode `\\12\catcode `\$12\catcode
  `\&12\catcode `\#12\catcode `\^12\catcode `\_12\catcode `\%12\relax}%
\providecommand \@@startlink[1]{}%
\providecommand \@@endlink[0]{}%
\providecommand \url  [0]{\begingroup\@sanitize@url \@url }%
\providecommand \@url [1]{\endgroup\@href {#1}{\urlprefix }}%
\providecommand \urlprefix  [0]{URL }%
\providecommand \Eprint [0]{\href }%
\providecommand \doibase [0]{http://dx.doi.org/}%
\providecommand \selectlanguage [0]{\@gobble}%
\providecommand \bibinfo  [0]{\@secondoftwo}%
\providecommand \bibfield  [0]{\@secondoftwo}%
\providecommand \translation [1]{[#1]}%
\providecommand \BibitemOpen [0]{}%
\providecommand \bibitemStop [0]{}%
\providecommand \bibitemNoStop [0]{.\EOS\space}%
\providecommand \EOS [0]{\spacefactor3000\relax}%
\providecommand \BibitemShut  [1]{\csname bibitem#1\endcsname}%
\let\auto@bib@innerbib\@empty
\bibitem [{\citenamefont {Yamada}\ \emph {et~al.}(2010)\citenamefont {Yamada},
  \citenamefont {Kulsrud},\ and\ \citenamefont {Ji}}]{YamadaRMP2010}%
  \BibitemOpen
  \bibfield  {author} {\bibinfo {author} {\bibfnamefont {M.}~\bibnamefont
  {Yamada}}, \bibinfo {author} {\bibfnamefont {R.}~\bibnamefont {Kulsrud}}, \
  and\ \bibinfo {author} {\bibfnamefont {H.}~\bibnamefont {Ji}},\ }\href
  {\doibase 10.1103/revmodphys.82.603} {\bibfield  {journal} {\bibinfo
  {journal} {Rev.\ Mod.\ Phys.}\ }\textbf {\bibinfo {volume} {82}},\
  \bibinfo {pages} {603} (\bibinfo {year} {2010})}\BibitemShut {NoStop}%
\bibitem [{\citenamefont {Kulsrud}(2005)}]{Kulsrud2005}%
  \BibitemOpen
  \bibfield  {author} {\bibinfo {author} {\bibfnamefont {R.}~\bibnamefont
  {Kulsrud}},\ }\href@noop {} {\emph {\bibinfo {title} {{Plasma Physics for
  Astrophysics}}}}\ (\bibinfo  {publisher} {Princeton University Press},\
  \bibinfo {address} {Princeton},\ \bibinfo {year} {2005})\BibitemShut
  {NoStop}%
\bibitem [{\citenamefont {Ji}\ and\ \citenamefont
  {Daughton}(2011)}]{JiPoP2011}%
  \BibitemOpen
  \bibfield  {author} {\bibinfo {author} {\bibfnamefont {H.}~\bibnamefont
  {Ji}}\ and\ \bibinfo {author} {\bibfnamefont {W.}~\bibnamefont {Daughton}},\
  }\href {\doibase 10.1063/1.3647505} {\bibfield  {journal} {\bibinfo
  {journal} {Phys.\ Plasmas}\ }\textbf {\bibinfo {volume} {18}},\ \bibinfo
  {pages} {111207} (\bibinfo {year} {2011})}\BibitemShut {NoStop}%
\bibitem [{\citenamefont {Nilson}\ \emph {et~al.}(2006)\citenamefont {Nilson},
  \citenamefont {Willingale}, \citenamefont {Kaluza}, \citenamefont
  {Kamperidis}, \citenamefont {Minardi}, \citenamefont {Wei}, \citenamefont
  {Fernandes}, \citenamefont {Notley}, \citenamefont {Bandyopadhyay},
  \citenamefont {Sherlock}, \citenamefont {Kingham}, \citenamefont {Tatarakis},
  \citenamefont {Najmudin}, \citenamefont {Rozmus}, \citenamefont {Evans},
  \citenamefont {Haines}, \citenamefont {Dangor},\ and\ \citenamefont
  {Krushelnick}}]{NilsonPRL2006}%
  \BibitemOpen
  \bibfield  {author} {\bibinfo {author} {\bibfnamefont {P.~M.}\ \bibnamefont
  {Nilson}}, \bibinfo {author} {\bibfnamefont {L.}~\bibnamefont {Willingale}},
  \bibinfo {author} {\bibfnamefont {M.~C.}\ \bibnamefont {Kaluza}}, \textit{et al.},\ }\href {\doibase
  10.1103/physrevlett.97.255001} {\bibfield  {journal} {\bibinfo  {journal}
  {Phys.\ Rev.\ Lett.}\ }\textbf {\bibinfo {volume} {97}},\ \bibinfo
  {pages} {255001} (\bibinfo {year} {2006})}\BibitemShut {NoStop}%
\bibitem [{\citenamefont {Li}\ \emph {et~al.}(2007)\citenamefont {Li},
  \citenamefont {S\'{e}guin}, \citenamefont {Frenje}, \citenamefont {Rygg},
  \citenamefont {Petrasso}, \citenamefont {Town}, \citenamefont {Landen},
  \citenamefont {Knauer},\ and\ \citenamefont {Smalyuk}}]{LiPRL2007b}%
  \BibitemOpen
  \bibfield  {author} {\bibinfo {author} {\bibfnamefont {C.~K.}\ \bibnamefont
  {Li}}, \bibinfo {author} {\bibfnamefont {F.~H.}\ \bibnamefont {S\'{e}guin}},
  \bibinfo {author} {\bibfnamefont {J.~A.}\ \bibnamefont {Frenje}}, \textit{et al.},\ }\href {\doibase
  10.1103/physrevlett.99.055001} {\bibfield  {journal} {\bibinfo  {journal}
  {Phys.\ Rev.\ Lett.}\ }\textbf {\bibinfo {volume} {99}},\ \bibinfo
  {pages} {055001} (\bibinfo {year} {2007})}\BibitemShut {NoStop}%
\bibitem [{\citenamefont {Rosenberg}\ \emph
  {et~al.}(2015{\natexlab{a}})\citenamefont {Rosenberg}, \citenamefont {Li},
  \citenamefont {Fox}, \citenamefont {Zylstra}, \citenamefont {Stoeckl},
  \citenamefont {S\'{e}guin}, \citenamefont {Frenje},\ and\ \citenamefont
  {Petrasso}}]{RosenbergPRL2015}%
  \BibitemOpen
  \bibfield  {author} {\bibinfo {author} {\bibfnamefont {M.~J.}\ \bibnamefont
  {Rosenberg}}, \bibinfo {author} {\bibfnamefont {C.~K.}\ \bibnamefont {Li}},
  \bibinfo {author} {\bibfnamefont {W.}~\bibnamefont {Fox}}, \textit{et al.},\ }\href {\doibase
  10.1103/physrevlett.114.205004} {\bibfield  {journal} {\bibinfo  {journal}
  {Phys. Rev. Lett.}\ }\textbf {\bibinfo {volume} {114}},\ \bibinfo {pages}
  {205004} (\bibinfo {year} {2015}{\natexlab{a}})}\BibitemShut {NoStop}%
\bibitem [{\citenamefont {Ren}\ \emph {et~al.}(2005)\citenamefont {Ren},
  \citenamefont {Yamada}, \citenamefont {Gerhardt}, \citenamefont {Ji},
  \citenamefont {Kulsrud},\ and\ \citenamefont {Kuritsyn}}]{RenPRL2005}%
  \BibitemOpen
  \bibfield  {author} {\bibinfo {author} {\bibfnamefont {Y.}~\bibnamefont
  {Ren}}, \bibinfo {author} {\bibfnamefont {M.}~\bibnamefont {Yamada}},
  \bibinfo {author} {\bibfnamefont {S.}~\bibnamefont {Gerhardt}}, \bibinfo
  {author} {\bibfnamefont {H.}~\bibnamefont {Ji}}, \bibinfo {author}
  {\bibfnamefont {R.}~\bibnamefont {Kulsrud}}, \ and\ \bibinfo {author}
  {\bibfnamefont {A.}~\bibnamefont {Kuritsyn}},\ }\href {\doibase
  10.1103/physrevlett.95.055003} {\bibfield  {journal} {\bibinfo  {journal}
  {Phys.\ Rev.\ Lett.}\ }\textbf {\bibinfo {volume} {95}},\ \bibinfo
  {pages} {055003} (\bibinfo {year} {2005})}\BibitemShut {NoStop}%
\bibitem [{\citenamefont {Fox}\ \emph {et~al.}(2017)\citenamefont {Fox},
  \citenamefont {Sciortino}, \citenamefont {Stechow}, \citenamefont
  {Jara-Almonte}, \citenamefont {Yoo}, \citenamefont {Ji},\ and\ \citenamefont
  {Yamada}}]{FoxPRL2017}%
  \BibitemOpen
  \bibfield  {author} {\bibinfo {author} {\bibfnamefont {W.}~\bibnamefont
  {Fox}}, \bibinfo {author} {\bibfnamefont {F.}~\bibnamefont {Sciortino}},
  \bibinfo {author} {\bibfnamefont {A.}~\bibnamefont {Stechow}}, \bibinfo
  {author} {\bibfnamefont {J.}~\bibnamefont {Jara-Almonte}}, \bibinfo {author}
  {\bibfnamefont {J.}~\bibnamefont {Yoo}}, \bibinfo {author} {\bibfnamefont
  {H.}~\bibnamefont {Ji}}, \ and\ \bibinfo {author} {\bibfnamefont
  {M.}~\bibnamefont {Yamada}},\ }\href {\doibase
  10.1103/physrevlett.118.125002} {\bibfield  {journal} {\bibinfo  {journal}
  {Phys.\ Rev.\ Lett.}\ }\textbf {\bibinfo {volume} {118}},\ \bibinfo
  {pages} {125002} (\bibinfo {year} {2017})}\BibitemShut {NoStop}%
\bibitem [{\citenamefont {Quataert}(2003)}]{QuataertAN2003}%
  \BibitemOpen
  \bibfield  {author} {\bibinfo {author} {\bibfnamefont {E.}~\bibnamefont
  {Quataert}},\ }\href {\doibase 10.1002/asna.200385043} {\bibfield  {journal}
  {\bibinfo  {journal} {Astron. Nachr.}\ }\textbf {\bibinfo {volume}
  {324}},\ \bibinfo {pages} {435} (\bibinfo {year} {2003})}\BibitemShut
  {NoStop}%
\bibitem [{\citenamefont {Matsumoto}\ \emph {et~al.}(2015)\citenamefont
  {Matsumoto}, \citenamefont {Amano}, \citenamefont {Kato},\ and\ \citenamefont
  {Hoshino}}]{MatsumotoScience2015}%
  \BibitemOpen
  \bibfield  {author} {\bibinfo {author} {\bibfnamefont {Y.}~\bibnamefont
  {Matsumoto}}, \bibinfo {author} {\bibfnamefont {T.}~\bibnamefont {Amano}},
  \bibinfo {author} {\bibfnamefont {T.~N.}\ \bibnamefont {Kato}}, \ and\
  \bibinfo {author} {\bibfnamefont {M.}~\bibnamefont {Hoshino}},\ }\href
  {\doibase 10.1126/science.1260168} {\bibfield  {journal} {\bibinfo  {journal}
  {Science}\ }\textbf {\bibinfo {volume} {347}},\ \bibinfo {pages} {974}
  (\bibinfo {year} {2015})}\BibitemShut {NoStop}%
\bibitem [{\citenamefont {Burch}\ \emph {et~al.}(2016)\citenamefont {Burch},
  \citenamefont {Torbert}, \citenamefont {Phan}, \citenamefont {Chen},
  \citenamefont {Moore}, \citenamefont {Ergun}, \citenamefont {Eastwood},
  \citenamefont {Gershman}, \citenamefont {Cassak}, \citenamefont {Argall},
  \citenamefont {Wang}, \citenamefont {Hesse}, \citenamefont {Pollock},
  \citenamefont {Giles}, \citenamefont {Nakamura}, \citenamefont {Mauk},
  \citenamefont {Fuselier}, \citenamefont {Russell}, \citenamefont
  {Strangeway}, \citenamefont {Drake}, \citenamefont {Shay}, \citenamefont
  {Khotyaintsev}, \citenamefont {Lindqvist}, \citenamefont {Marklund},
  \citenamefont {Wilder}, \citenamefont {Young}, \citenamefont {Torkar},
  \citenamefont {Goldstein}, \citenamefont {Dorelli}, \citenamefont {Avanov},
  \citenamefont {Oka}, \citenamefont {Baker}, \citenamefont {Jaynes},
  \citenamefont {Goodrich}, \citenamefont {Cohen}, \citenamefont {Turner},
  \citenamefont {Fennell}, \citenamefont {Blake}, \citenamefont {Clemmons},
  \citenamefont {Goldman}, \citenamefont {Newman}, \citenamefont {Petrinec},
  \citenamefont {Trattner}, \citenamefont {Lavraud}, \citenamefont {Reiff},
  \citenamefont {Baumjohann}, \citenamefont {Magnes}, \citenamefont {Steller},
  \citenamefont {Lewis}, \citenamefont {Saito}, \citenamefont {Coffey},\ and\
  \citenamefont {Chandler}}]{BurchScience2016}%
  \BibitemOpen
  \bibfield  {author} {\bibinfo {author} {\bibfnamefont {J.~L.}\ \bibnamefont
  {Burch}}, \bibinfo {author} {\bibfnamefont {R.~B.}\ \bibnamefont {Torbert}},
  \bibinfo {author} {\bibfnamefont {T.~D.}\ \bibnamefont {Phan}}, \textit{et al.},\ }\href {\doibase 10.1126/science.aaf2939}
  {\bibfield  {journal} {\bibinfo  {journal} {Science}\ }\textbf {\bibinfo
  {volume} {352}},\ \bibinfo {pages} {aaf2939} (\bibinfo {year}
  {2016})}\BibitemShut {NoStop}%
\bibitem [{\citenamefont {Loureiro}\ \emph {et~al.}(2007)\citenamefont
  {Loureiro}, \citenamefont {Schekochihin},\ and\ \citenamefont
  {Cowley}}]{LoureiroPoP2007}%
  \BibitemOpen
  \bibfield  {author} {\bibinfo {author} {\bibfnamefont {N.~F.}\ \bibnamefont
  {Loureiro}}, \bibinfo {author} {\bibfnamefont {A.~A.}\ \bibnamefont
  {Schekochihin}}, \ and\ \bibinfo {author} {\bibfnamefont {S.~C.}\
  \bibnamefont {Cowley}},\ }\href {\doibase 10.1063/1.2783986} {\bibfield
  {journal} {\bibinfo  {journal} {Phys.\ Plasmas}\ }\textbf {\bibinfo
  {volume} {14}},\ \bibinfo {pages} {100703} (\bibinfo {year}
  {2007})}\BibitemShut {NoStop}%
\bibitem [{\citenamefont {Bhattacharjee}\ \emph {et~al.}(2009)\citenamefont
  {Bhattacharjee}, \citenamefont {Huang}, \citenamefont {Yang},\ and\
  \citenamefont {Rogers}}]{BhattacharjeePoP2009}%
  \BibitemOpen
  \bibfield  {author} {\bibinfo {author} {\bibfnamefont {A.}~\bibnamefont
  {Bhattacharjee}}, \bibinfo {author} {\bibfnamefont {Y.~M.}\ \bibnamefont
  {Huang}}, \bibinfo {author} {\bibfnamefont {H.}~\bibnamefont {Yang}}, \ and\
  \bibinfo {author} {\bibfnamefont {B.}~\bibnamefont {Rogers}},\ }\href
  {\doibase 10.1063/1.3264103} {\bibfield  {journal} {\bibinfo  {journal}
  {Phys.\ Plasmas}\ }\textbf {\bibinfo {volume} {16}},\ \bibinfo {pages}
  {112102} (\bibinfo {year} {2009})}\BibitemShut {NoStop}%
\bibitem [{\citenamefont {Cai}\ and\ \citenamefont {Lee}(1997)}]{CaiPoP1997}%
  \BibitemOpen
  \bibfield  {author} {\bibinfo {author} {\bibfnamefont {H.~J.}\ \bibnamefont
  {Cai}}\ and\ \bibinfo {author} {\bibfnamefont {L.~C.}\ \bibnamefont {Lee}},\
  }\href {\doibase 10.1063/1.872178} {\bibfield  {journal} {\bibinfo  {journal}
  {Phys.\ Plasmas}\ }\textbf {\bibinfo {volume} {4}},\ \bibinfo {pages}
  {509} (\bibinfo {year} {1997})}\BibitemShut {NoStop}%
\bibitem [{\citenamefont {Hesse}\ \emph {et~al.}(1999)\citenamefont {Hesse},
  \citenamefont {Schindler}, \citenamefont {Birn},\ and\ \citenamefont
  {Kuznetsova}}]{HessePoP1999}%
  \BibitemOpen
  \bibfield  {author} {\bibinfo {author} {\bibfnamefont {M.}~\bibnamefont
  {Hesse}}, \bibinfo {author} {\bibfnamefont {K.}~\bibnamefont {Schindler}},
  \bibinfo {author} {\bibfnamefont {J.}~\bibnamefont {Birn}}, \ and\ \bibinfo
  {author} {\bibfnamefont {M.}~\bibnamefont {Kuznetsova}},\ }\href {\doibase
  10.1063/1.873436} {\bibfield  {journal} {\bibinfo  {journal} {Phys.
  Plasmas}\ }\textbf {\bibinfo {volume} {6}},\ \bibinfo {pages} {1781}
  (\bibinfo {year} {1999})}\BibitemShut {NoStop}%
\bibitem [{\citenamefont {Ji}\ \emph {et~al.}(2008)\citenamefont {Ji},
  \citenamefont {Ren}, \citenamefont {Yamada}, \citenamefont {Dorfman},
  \citenamefont {Daughton},\ and\ \citenamefont {Gerhardt}}]{JiGRL2008}%
  \BibitemOpen
  \bibfield  {author} {\bibinfo {author} {\bibfnamefont {H.}~\bibnamefont
  {Ji}}, \bibinfo {author} {\bibfnamefont {Y.}~\bibnamefont {Ren}}, \bibinfo
  {author} {\bibfnamefont {M.}~\bibnamefont {Yamada}}, \textit{et al.},\ }\href {\doibase 10.1029/2008gl034538}
  {\bibfield  {journal} {\bibinfo  {journal} {Geophys.\ Res.\ Lett.}\
  }\textbf {\bibinfo {volume} {35}},\ \bibinfo {pages} {L13106} (\bibinfo
  {year} {2008})}\BibitemShut {NoStop}%
\bibitem [{\citenamefont {Phan}\ \emph {et~al.}(2018)\citenamefont {Phan},
  \citenamefont {Eastwood}, \citenamefont {Shay}, \citenamefont {Drake},
  \citenamefont {Sonnerup}, \citenamefont {Fujimoto}, \citenamefont {Cassak},
  \citenamefont {{\O}ieroset}, \citenamefont {Burch}, \citenamefont {Torbert},
  \citenamefont {Rager}, \citenamefont {Dorelli}, \citenamefont {Gershman},
  \citenamefont {Pollock}, \citenamefont {Pyakurel}, \citenamefont {Haggerty},
  \citenamefont {Khotyaintsev}, \citenamefont {Lavraud}, \citenamefont {Saito},
  \citenamefont {Oka}, \citenamefont {Ergun}, \citenamefont {Retino},
  \citenamefont {{Le Contel}}, \citenamefont {Argall}, \citenamefont {Giles},
  \citenamefont {Moore}, \citenamefont {Wilder}, \citenamefont {Strangeway},
  \citenamefont {Russell}, \citenamefont {Lindqvist},\ and\ \citenamefont
  {Magnes}}]{PhanNature2018}%
  \BibitemOpen
  \bibfield  {author} {\bibinfo {author} {\bibfnamefont {T.~D.}\ \bibnamefont
  {Phan}}, \bibinfo {author} {\bibfnamefont {J.~P.}\ \bibnamefont {Eastwood}},
  \bibinfo {author} {\bibfnamefont {M.~A.}\ \bibnamefont {Shay}}, \textit{et al.},\ }\href {\doibase
  10.1038/s41586-018-0091-5} {\bibfield  {journal} {\bibinfo  {journal}
  {Nature}\ }\textbf {\bibinfo {volume} {557}},\ \bibinfo {pages} {202}
  (\bibinfo {year} {2018})}\BibitemShut {NoStop}%
\bibitem [{\citenamefont {Yates}\ \emph {et~al.}(1982)\citenamefont {Yates},
  \citenamefont {van Hulsteyn}, \citenamefont {Rutkowski}, \citenamefont
  {Kyrala},\ and\ \citenamefont {Brackbill}}]{YatesPRL1982}%
  \BibitemOpen
  \bibfield  {author} {\bibinfo {author} {\bibfnamefont {M.~A.}\ \bibnamefont
  {Yates}}, \bibinfo {author} {\bibfnamefont {D.~B.}\ \bibnamefont {van
  Hulsteyn}}, \bibinfo {author} {\bibfnamefont {H.}~\bibnamefont {Rutkowski}},\textit{et al.},\ }\href {\doibase
  10.1103/physrevlett.49.1702} {\bibfield  {journal} {\bibinfo  {journal}
  {Phys.\ Rev.\ Lett.}\ }\textbf {\bibinfo {volume} {49}},\ \bibinfo
  {pages} {1702} (\bibinfo {year} {1982})}\BibitemShut {NoStop}%
\bibitem [{\citenamefont {Matteucci}\ \emph {et~al.}(2018)\citenamefont
  {Matteucci}, \citenamefont {Fox}, \citenamefont {Bhattacharjee},
  \citenamefont {Schaeffer}, \citenamefont {Moissard}, \citenamefont
  {Germaschewski}, \citenamefont {Fiksel},\ and\ \citenamefont
  {Hu}}]{MatteucciPRL2018}%
  \BibitemOpen
  \bibfield  {author} {\bibinfo {author} {\bibfnamefont {J.}~\bibnamefont
  {Matteucci}}, \bibinfo {author} {\bibfnamefont {W.}~\bibnamefont {Fox}},
  \bibinfo {author} {\bibfnamefont {A.}~\bibnamefont {Bhattacharjee}}, \textit{et al.},\ }\href@noop {} {\bibfield
  {journal} {\bibinfo  {journal} {Phys. Rev. Lett.}\ }\textbf {\bibinfo
  {volume} {121}},\ \bibinfo {pages} {095001} (\bibinfo {year}
  {2018})}\BibitemShut {NoStop}%
\bibitem [{\citenamefont {Fiksel}\ \emph {et~al.}(2014)\citenamefont {Fiksel},
  \citenamefont {Fox}, \citenamefont {Bhattacharjee}, \citenamefont {Barnak},
  \citenamefont {Chang}, \citenamefont {Germaschewski}, \citenamefont {Hu},\
  and\ \citenamefont {Nilson}}]{FikselPRL2014}%
  \BibitemOpen
  \bibfield  {author} {\bibinfo {author} {\bibfnamefont {G.}~\bibnamefont
  {Fiksel}}, \bibinfo {author} {\bibfnamefont {W.}~\bibnamefont {Fox}},
  \bibinfo {author} {\bibfnamefont {A.}~\bibnamefont {Bhattacharjee}},  \textit{et al.},\ }\href {\doibase
  10.1103/physrevlett.113.105003} {\bibfield  {journal} {\bibinfo  {journal}
  {Phys.\ Rev.\ Lett.}\ }\textbf {\bibinfo {volume} {113}},\ \bibinfo
  {pages} {105003} (\bibinfo {year} {2014})}\BibitemShut {NoStop}%
\bibitem [{\citenamefont {Rosenberg}\ \emph
  {et~al.}(2015{\natexlab{b}})\citenamefont {Rosenberg}, \citenamefont {Li},
  \citenamefont {Fox}, \citenamefont {Igumenshchev}, \citenamefont
  {S\'{e}guin}, \citenamefont {Town}, \citenamefont {Frenje}, \citenamefont
  {Stoeckl}, \citenamefont {Glebov},\ and\ \citenamefont
  {Petrasso}}]{RosenbergNatComm2015}%
  \BibitemOpen
  \bibfield  {author} {\bibinfo {author} {\bibfnamefont {M.~J.}\ \bibnamefont
  {Rosenberg}}, \bibinfo {author} {\bibfnamefont {C.~K.}\ \bibnamefont {Li}},
  \bibinfo {author} {\bibfnamefont {W.}~\bibnamefont {Fox}}, \textit{et al.},\ }\href {\doibase 10.1038/ncomms7190} {\bibfield
  {journal} {\bibinfo  {journal} {Nature Comms.}\ }\textbf {\bibinfo
  {volume} {6}},\ \bibinfo {pages} {6190} (\bibinfo {year}
  {2015}{\natexlab{b}})}\BibitemShut {NoStop}%
\bibitem [{\citenamefont {Oertel}\ \emph {et~al.}(2006)\citenamefont {Oertel},
  \citenamefont {Aragonez}, \citenamefont {Archuleta}, \citenamefont {Barnes},
  \citenamefont {Casper}, \citenamefont {Fatherley}, \citenamefont {Heinrichs},
  \citenamefont {King}, \citenamefont {Landers}, \citenamefont {Lopez},
  \citenamefont {Sanchez}, \citenamefont {Sandoval}, \citenamefont {Schrank},
  \citenamefont {Walsh}, \citenamefont {Bell}, \citenamefont {Brown},
  \citenamefont {Costa}, \citenamefont {Holder}, \citenamefont {Montelongo},\
  and\ \citenamefont {Pederson}}]{OertelRSI2006}%
  \BibitemOpen
  \bibfield  {author} {\bibinfo {author} {\bibfnamefont {J.~A.}\ \bibnamefont
  {Oertel}}, \bibinfo {author} {\bibfnamefont {R.}~\bibnamefont {Aragonez}},
  \bibinfo {author} {\bibfnamefont {T.}~\bibnamefont {Archuleta}}, \textit{et al.},\ }\href {\doibase
  10.1063/1.2227439} {\bibfield  {journal} {\bibinfo  {journal} {Rev. 
  Sci. Instrum.}\ }\textbf {\bibinfo {volume} {77}},\ \bibinfo {pages}
  {10E308} (\bibinfo {year} {2006})}\BibitemShut {NoStop}%
\bibitem [{\citenamefont {S{\'{e}}guin}\ \emph {et~al.}(2003)\citenamefont
  {S{\'{e}}guin}, \citenamefont {Frenje}, \citenamefont {Li}, \citenamefont
  {Hicks}, \citenamefont {Kurebayashi}, \citenamefont {Rygg}, \citenamefont
  {Schwartz}, \citenamefont {Petrasso}, \citenamefont {Roberts}, \citenamefont
  {Soures}, \citenamefont {Meyerhofer}, \citenamefont {Sangster}, \citenamefont
  {Knauer}, \citenamefont {Sorce}, \citenamefont {Glebov}, \citenamefont
  {Stoeckl}, \citenamefont {Phillips}, \citenamefont {Leeper}, \citenamefont
  {Fletcher},\ and\ \citenamefont {Padalino}}]{SeguinRSI2003}%
  \BibitemOpen
  \bibfield  {author} {\bibinfo {author} {\bibfnamefont {F.~H.}\ \bibnamefont
  {S{\'{e}}guin}}, \bibinfo {author} {\bibfnamefont {J.~A.}\ \bibnamefont
  {Frenje}}, \bibinfo {author} {\bibfnamefont {C.~K.}\ \bibnamefont {Li}}, \textit{et al.},\ }\href {\doibase
  10.1063/1.1518141} {\bibfield  {journal} {\bibinfo  {journal} {Rev.
  Sci. Instrum.}\ }\textbf {\bibinfo {volume} {74}},\ \bibinfo {pages}
  {975} (\bibinfo {year} {2003})}\BibitemShut {NoStop}%
\bibitem [{\citenamefont {Petrasso}\ \emph {et~al.}(2009)\citenamefont
  {Petrasso}, \citenamefont {Li}, \citenamefont {Seguin}, \citenamefont {Rygg},
  \citenamefont {Frenje}, \citenamefont {Betti}, \citenamefont {Knauer},
  \citenamefont {Meyerhofer}, \citenamefont {Amendt}, \citenamefont {Froula},
  \citenamefont {Landen}, \citenamefont {Patel}, \citenamefont {Ross},\ and\
  \citenamefont {Town}}]{PetrassoPRL2009}%
  \BibitemOpen
  \bibfield  {author} {\bibinfo {author} {\bibfnamefont {R.~D.}\ \bibnamefont
  {Petrasso}}, \bibinfo {author} {\bibfnamefont {C.~K.}\ \bibnamefont {Li}},
  \bibinfo {author} {\bibfnamefont {F.~H.}\ \bibnamefont {Seguin}}, \textit{et al.},\ }\href {\doibase 10.1103/physrevlett.103.085001} {\bibfield
  {journal} {\bibinfo  {journal} {Phys. Rev. Lett.}\ }\textbf {\bibinfo
  {volume} {103}},\ \bibinfo {pages} {085001} (\bibinfo {year}
  {2009})}\BibitemShut {NoStop}%
\bibitem [{\citenamefont {Bott}\ \emph {et~al.}(2017)\citenamefont {Bott},
  \citenamefont {Graziani}, \citenamefont {Tzeferacos}, \citenamefont {White},
  \citenamefont {Lamb}, \citenamefont {Gregori},\ and\ \citenamefont
  {Schekochihin}}]{BottJPP2017}%
  \BibitemOpen
  \bibfield  {author} {\bibinfo {author} {\bibfnamefont {A.~F.~A.}\
  \bibnamefont {Bott}}, \bibinfo {author} {\bibfnamefont {C.}~\bibnamefont
  {Graziani}}, \bibinfo {author} {\bibfnamefont {P.}~\bibnamefont
  {Tzeferacos}}, \textit{et al.},\  }\href {\doibase 10.1017/s0022377817000939} {\bibfield  {journal} {\bibinfo
  {journal} {J. Plasma Phys.}\ }\textbf {\bibinfo {volume} {83}}
  (\bibinfo {year} {2017})}\BibitemShut {NoStop}%
\bibitem [{Note1()}]{Note1}%
  \BibitemOpen
  \bibinfo {note} {See extended material}\BibitemShut {NoStop}%
\bibitem [{\citenamefont {S{\'e}guin}\ \emph {et~al.}(2012)\citenamefont
  {S{\'e}guin}, \citenamefont {Li}, \citenamefont {Manuel}, \citenamefont
  {Rinderknecht}, \citenamefont {Sinenian}, \citenamefont {Frenje},
  \citenamefont {Rygg}, \citenamefont {Hicks}, \citenamefont {Petrasso},
  \citenamefont {Delettrez}, \citenamefont {Betti}, \citenamefont {Marshall},\
  and\ \citenamefont {Smalyuk}}]{SeguinPoP2012}%
  \BibitemOpen
  \bibfield  {author} {\bibinfo {author} {\bibfnamefont {F.~H.}\ \bibnamefont
  {S{\'e}guin}}, \bibinfo {author} {\bibfnamefont {C.~K.}\ \bibnamefont {Li}},
  \bibinfo {author} {\bibfnamefont {M.~J.-E.}\ \bibnamefont {Manuel}}, \textit{et al.},\ }\href {\doibase 10.1063/1.3671908}
  {\bibfield  {journal} {\bibinfo  {journal} {Phys.\ Plasmas}\ }\textbf
  {\bibinfo {volume} {19}},\ \bibinfo {pages} {012701} (\bibinfo {year}
  {2012})}\BibitemShut {NoStop}%
\bibitem [{\citenamefont {Fox}\ \emph {et~al.}(2018)\citenamefont {Fox},
  \citenamefont {Matteucci}, \citenamefont {Moissard}, \citenamefont
  {Schaeffer}, \citenamefont {Bhattacharjee}, \citenamefont {Germaschewski},\
  and\ \citenamefont {Hu}}]{FoxPoP2018}%
  \BibitemOpen
  \bibfield  {author} {\bibinfo {author} {\bibfnamefont {W.}~\bibnamefont
  {Fox}}, \bibinfo {author} {\bibfnamefont {J.}~\bibnamefont {Matteucci}},
  \bibinfo {author} {\bibfnamefont {C.}~\bibnamefont {Moissard}}, \textit{et al.}\ }\href {\doibase
  10.1063/1.5050813} {\bibfield  {journal} {\bibinfo  {journal} {Phys.
  Plasmas}\ }\textbf {\bibinfo {volume} {25}},\ \bibinfo {pages} {102106}
  (\bibinfo {year} {2018})}\BibitemShut {NoStop}%
\bibitem [{Note2()}]{Note2}%
  \BibitemOpen
  \bibinfo {note} {PSC simulations matched to the plasma conditions showed that
  Biermann battery magnetic field generation was nearly equal between the
  inboard and outboard sides}\BibitemShut {NoStop}%
\bibitem [{\citenamefont {Kuritsyn}\ \emph {et~al.}(2006)\citenamefont
  {Kuritsyn}, \citenamefont {Yamada}, \citenamefont {Gerhardt}, \citenamefont
  {Ji}, \citenamefont {Kulsrud},\ and\ \citenamefont {Ren}}]{KuritsynPoP2006}%
  \BibitemOpen
  \bibfield  {author} {\bibinfo {author} {\bibfnamefont {A.}~\bibnamefont
  {Kuritsyn}}, \bibinfo {author} {\bibfnamefont {M.}~\bibnamefont {Yamada}},
  \bibinfo {author} {\bibfnamefont {S.}~\bibnamefont {Gerhardt}}, \bibinfo
  {author} {\bibfnamefont {H.}~\bibnamefont {Ji}}, \bibinfo {author}
  {\bibfnamefont {R.}~\bibnamefont {Kulsrud}}, \ and\ \bibinfo {author}
  {\bibfnamefont {Y.}~\bibnamefont {Ren}},\ }\href {\doibase 10.1063/1.2179416}
  {\bibfield  {journal} {\bibinfo  {journal} {Phys.\ Plasmas}\ }\textbf
  {\bibinfo {volume} {13}},\ \bibinfo {pages} {055703} (\bibinfo {year}
  {2006})}\BibitemShut {NoStop}%
\bibitem [{\citenamefont {Lezhnin}\ \emph {et~al.}(2018)\citenamefont
  {Lezhnin}, \citenamefont {Fox}, \citenamefont {Matteucci}, \citenamefont
  {Schaeffer}, \citenamefont {Bhattacharjee}, \citenamefont {Rosenberg},\ and\
  \citenamefont {Germaschewski}}]{LezhninPoP2018}%
  \BibitemOpen
  \bibfield  {author} {\bibinfo {author} {\bibfnamefont {K.~V.}\ \bibnamefont
  {Lezhnin}}, \bibinfo {author} {\bibfnamefont {W.}~\bibnamefont {Fox}},
  \bibinfo {author} {\bibfnamefont {J.}~\bibnamefont {Matteucci}}, \bibinfo
  {author} {\bibfnamefont {D.~B.}\ \bibnamefont {Schaeffer}}, \bibinfo {author}
  {\bibfnamefont {A.}~\bibnamefont {Bhattacharjee}}, \bibinfo {author}
  {\bibfnamefont {M.~J.}\ \bibnamefont {Rosenberg}}, \ and\ \bibinfo {author}
  {\bibfnamefont {K.}~\bibnamefont {Germaschewski}},\ }\href {\doibase
  10.1063/1.5044547} {\bibfield  {journal} {\bibinfo  {journal} {Phys.
  Plasmas}\ }\textbf {\bibinfo {volume} {25}},\ \bibinfo {pages} {093105}
  (\bibinfo {year} {2018})}\BibitemShut {NoStop}%
\bibitem [{\citenamefont {Fox}\ \emph {et~al.}(2011)\citenamefont {Fox},
  \citenamefont {Bhattacharjee},\ and\ \citenamefont
  {Germaschewski}}]{FoxPRL2011}%
  \BibitemOpen
  \bibfield  {author} {\bibinfo {author} {\bibfnamefont {W.}~\bibnamefont
  {Fox}}, \bibinfo {author} {\bibfnamefont {A.}~\bibnamefont {Bhattacharjee}},
  \ and\ \bibinfo {author} {\bibfnamefont {K.}~\bibnamefont {Germaschewski}},\
  }\href {\doibase 10.1103/physrevlett.106.215003} {\bibfield  {journal}
  {\bibinfo  {journal} {Phys.\ Rev.\ Lett.}\ }\textbf {\bibinfo {volume}
  {106}},\ \bibinfo {pages} {215003} (\bibinfo {year} {2011})}\BibitemShut
  {NoStop}%
\bibitem [{\citenamefont {Matteucci}\ \emph {et~al.}(2019)\citenamefont
  {Matteucci}, \citenamefont {Fox}, \citenamefont {Bhattacharjee},
  \citenamefont {Schaeffer}, \citenamefont {Moissard}, \citenamefont
  {Germaschewski}, \citenamefont {Fiksel},\ and\ \citenamefont
  {Hu}}]{MatteucciPoP2019}%
  \BibitemOpen
  \bibfield  {author} {\bibinfo {author} {\bibfnamefont {J.}~\bibnamefont
  {Matteucci}}, \bibinfo {author} {\bibfnamefont {W.}~\bibnamefont {Fox}},
  \bibinfo {author} {\bibfnamefont {A.}~\bibnamefont {Bhattacharjee}}, \bibinfo
  {author} {\bibfnamefont {D.~B.}\ \bibnamefont {Schaeffer}}, \bibinfo {author}
  {\bibfnamefont {C.}~\bibnamefont {Moissard}}, \bibinfo {author}
  {\bibfnamefont {K.}~\bibnamefont {Germaschewski}}, \bibinfo {author}
  {\bibfnamefont {G.}~\bibnamefont {Fiksel}}, \ and\ \bibinfo {author}
  {\bibfnamefont {S.~X.}\ \bibnamefont {Hu}},\ }\href@noop {} {\bibfield
  {journal} {\bibinfo  {journal} {to be submitted to Phys. Plasmas}\ }
  (\bibinfo {year} {2020})}\BibitemShut {NoStop}%
\bibitem [{\citenamefont {Kugland}\ \emph {et~al.}(2012)\citenamefont
  {Kugland}, \citenamefont {Ryutov}, \citenamefont {Plechaty}, \citenamefont
  {Ross},\ and\ \citenamefont {Park}}]{KuglandRSI2012}%
  \BibitemOpen
  \bibfield  {author} {\bibinfo {author} {\bibfnamefont {N.~L.}\ \bibnamefont
  {Kugland}}, \bibinfo {author} {\bibfnamefont {D.~D.}\ \bibnamefont {Ryutov}},
  \bibinfo {author} {\bibfnamefont {C.}~\bibnamefont {Plechaty}}, \bibinfo
  {author} {\bibfnamefont {J.~S.}\ \bibnamefont {Ross}}, \ and\ \bibinfo
  {author} {\bibfnamefont {H.~S.}\ \bibnamefont {Park}},\ }\href {\doibase
  10.1063/1.4750234} {\bibfield  {journal} {\bibinfo  {journal} {Rev.
  Sci. Instrum.}\ }\textbf {\bibinfo {volume} {83}},\ \bibinfo {pages}
  {101301} (\bibinfo {year} {2012})}\BibitemShut {NoStop}%
\bibitem [{\citenamefont {Kasim}\ \emph {et~al.}(2019)\citenamefont {Kasim},
  \citenamefont {Bott}, \citenamefont {Tzeferacos}, \citenamefont {Lamb},
  \citenamefont {Gregori},\ and\ \citenamefont {Vinko}}]{KasimPRE2019}%
  \BibitemOpen
  \bibfield  {author} {\bibinfo {author} {\bibfnamefont {M.~F.}\ \bibnamefont
  {Kasim}}, \bibinfo {author} {\bibfnamefont {A.~F.~A.}\ \bibnamefont {Bott}},
  \bibinfo {author} {\bibfnamefont {P.}~\bibnamefont {Tzeferacos}}, \bibinfo
  {author} {\bibfnamefont {D.~Q.}\ \bibnamefont {Lamb}}, \bibinfo {author}
  {\bibfnamefont {G.}~\bibnamefont {Gregori}}, \ and\ \bibinfo {author}
  {\bibfnamefont {S.~M.}\ \bibnamefont {Vinko}},\ }\href {\doibase
  10.1103/physreve.100.033208} {\bibfield  {journal} {\bibinfo  {journal}
  {Phys.\ Rev. E}\ }\textbf {\bibinfo {volume} {100}},\ \bibinfo {pages}
  {033208} (\bibinfo {year} {2019})}\BibitemShut {NoStop}%
\end{thebibliography}
%
%

%


\appendix
\clearpage

\section{1-D proton inversion algorithm: PRADICAMENT}

This section describes the fast 1-D inversion algorithm PRADICAMENT 
used for obtaining the 1-D magnetic field  from an observed proton fluence profile.
A manuscript describing PRADICAMENT and its benchmarking against
other inversion algorithms such as PROBLEM \cite{BottJPP2017} is in preparation.
A 1-D inversion routine has the utility that it can be implemented 
very simply and can run very fast, which is useful for uncertainly
analysis, and therefore confidence in the results.

We briefly review the proton radiography setup to fix the terminology and relevant
constants. 
For this analysis, consider probing a one-dimensional magnetic structure following the approach outlined in Ref.~\cite{KuglandRSI2012}.
We assume the probe beam propagates along the $z$-axis of a cartesian coordinate system,
and is deflected by the field $B_y(x)$ similar to the experiment geometry.
The protons emerge from a 
point-source located at a distance $L_1$ from the object.  The detector is 
positioned at a distance $L_2$ on the opposite side of the the object.  
We assume the proton beam is low-divergence and interaction with the magnetic field can be treated in a paraxial approximation.

If a proton crosses the plasma at a coordinate $x$, its position $X$ at the detector plane is \cite{KuglandRSI2012},
\begin{equation}
\label{}
X = (1+L_2/L_1) x + L_2 \frac{e}{mv_{p}}\int{B_y(x,z)dz},
\end{equation}
where $e$, $m$, and $v_{p}$ are the proton charge, mass, and velocity and the $B$-integral is taken along the beam propagation.
The first term in the  \textit{RHS} describes the ballistic  zero-field  propagation with a geometric magnification $M=(L_1+L_2)/L_1$ and the second term
describes deflection by the magnetic field.

It is convenient to work in the coordinate system of the object (plasma) plane, so we introduce
$x^{\prime}=X/M$, and use $\bar{B}(x)=\int{B_y(x,z)dz}$.  This equation can then be written in a simplified form,
\begin{equation}
\label{eq:deflection}
x^{\prime} = x+\nu \bar{B}(x),
\end{equation}
where $\nu=L_1 L_2 / (L_1+L_2) \cdot e/mv_{p}$.
We refer to $x^\prime$ as the ``final'' proton coordinate, in contrast to the ``initial'' 
coordinate $x$.

The proton flux intensity transforms according to the transformation of the linear dimension,
\begin{equation}
\label{eq:fluence}
I(x^\prime) = \frac{I_{0}(x)}{dx^{\prime}/dx},
\end{equation}
where $I(x^\prime)$ is the experimentally-measured 
proton fluence and $I_0(x)$ a specified undisturbed proton fluence at the object plane.
The subtlety is that the $I$ data is observed at the final coordinates,
but depends on the undisturbed fluence from the initial coordinates.
The user also specifies the incident proton fluence
profile $I_0(x)$, which can have spatial variation.

After substituting the derivatives in Eq.~\ref{eq:fluence} using Eq.~\ref{eq:deflection}, an expression for the magnetic field as a function of $I$ and $I_0$ is, 
\begin{equation}
\label{eq:Bderivative}
\frac{{d\bar B}}{{dx}} = {\nu ^{ - 1}}\left( {\frac{{{I_0(x)}}}{{I({x^\prime })}} - 1} \right).
\end{equation}
The simple form of Eq.~\ref{eq:Bderivative} is rather deceptive since the \textit{LHS} of the equation has the magnetic field
as a function of the initial proton coordinates $x$ while the \textit{RHS} depends
as well on the final proton coordinates $x^{\prime}$ through $I(x^\prime$), which are in turn coupled through Eq.~\ref{eq:deflection}.

The ODE,  Eq.~\ref{eq:Bderivative}, is straightforward to solve 
using standard Matlab ODE solvers.
Valid solutions exist provided that 
for each value of $x^\prime$ there exists one and only one value of $x$.
This assumption holds if the magnetic field is limited in magnitude to an extent that
 $dx^{\prime}/dx > 0$, which is equivalent to the absence of
caustics in the proton image as described in Ref.~\cite{KuglandRSI2012},
or equivalently that proton trajectories do not cross.

The algorithm begins at a chosen $x^\prime_0$ with initial $B=B_0$.
$B(x)$ is determined from that point 
outwards in both directions using the ODE solver.
As needed, $I(x^\prime)$ is interpolated 
between gridded points.
To check the solution, Eq.~\ref{eq:fluence} is 
directly evaluated from $B(x)$ and compared to the input $I(x^\prime)$.
The algorithm was also directly tested against
several analytic profiles.
 
We conclude with some notes about boundary conditions.
Given the structure of Eq.~\ref{eq:Bderivative}, 
the solution requires at least one boundary condition.  That is,
a constant $B$ can always be added to the 
solution, which results in only a spatial offset between $x$ and $x^\prime$.
This constant offset $B$ in general must be determined from known boundary conditions
or other constraints.

Second, the initial proton fluence
$I_0(x)$ is also in principle unknown.  (It is not known to the authors
how to separately measure this quantity along the same line of sight as $I(x^\prime)$.
Ref.~\cite{KasimPRE2019} discusses another recent approach
to constraining $B$-field profiles without direct knowledge of the undisturbed fluence.)
Assuming $I_0$ is a constant or smoothly
varying profile, the average fluence and any
parameters of the profile must be constrained 
as part of the analysis procedure.
We find that multiplying $I_0$ by
a constant factor (i.e. assuming a different average value)
approximately adds 
a linear component $B_1 \approx \alpha x$ to $B(x)$.
Therefore, constraining the average proton
fluence requires an additional boundary condition on $B$.
By similar argument, adding higher-order structure to $I_0$, such as 
parabolicity, adds corresponding higher-order components
to $B$, and in general for each
free parameter added to $I_0$ an additional constraint is needed.

\end{document}